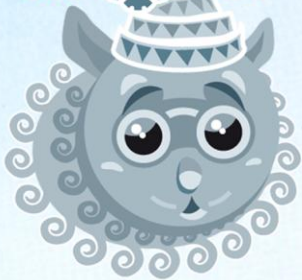

**НТ 37**
**КАЗАНЬ - 2015**

Российская академия наук
Отделение физических наук РАН
Научный совет РАН по физике низких температур
Академия наук Республики Татарстан
Институт физических проблем им. П.Л. Капицы РАН
Казанский федеральный университет

# XXXVII Совещание по физике низких температур

## Программа, тезисы докладов

Казань
29 июня - 3 июля 2015

# Photoconduction and Low-Temperature Ohmic Conduction of Peierls Conductor o-TaS$_3$ under Uniaxial Strain

V.E. Minakova[1], A.N. Taldenkov[2], S.V. Zaitzev-Zotov[1]

[1]Kotel'nikov Institute of Radio Engineering & Electronics of RAS, 125009, Mokhovaya 11-7, Moscow, Russia

[2]National Research Centre «Kurchatov Institute», 123182, 1, Akademika Kurchatova pl., Moscow, Russia

e-mail: mina_cplire@mail.ru

It is well known [1] that in quasi-one-dimensional conductor orthorhombic TaS$_3$ ($o$-TaS$_3$) below the Peierls transition temperature, $T_P \approx 220$ K, all conduction electrons are condensed into a charge-density-wave (CDW) state and at low electric field, $E$, do not contribute to the conductance, $G(T)$, provided by quasi-particles thermally exited over the Peierls gap and obeying an activation law with the activation energy $E_\Delta \approx 800$ K (Ohmic conductance). $G(T)$ becomes strongly non-linear at $E > E_T$ ($E_T$ – the threshold field for CDW depinning) due to CDW sliding, which is accompanied by generation of narrow-band-noise (NBN), whose frequency is proportional to CDW velocity. Below $T \leq T_P/2$ the Ohmic conductance in the chain direction begins to deviate from the initial activation law, a new activation energy, $E_L$, being approximately half [2], while the perpendicular conductance preserves the initial value $E_{\Delta\perp} \approx 800$ K in all temperature range. A transition to the new activation law is often accompanied by an appearance of a plateau with a weakly dependent conductance connecting the different activation parts of $G(T)$-curve. The nature of the low-temperature Ohmic conduction is attributed to collective excitations of the CDW, presumably solitons [2, 3].

At high $T$ the CDW wave vector, $q$, is slightly incommensurate with the lattice one and tends to commensurability when $T$ decreases to T $\approx 30$ K [4]. A strain, $\varepsilon$, applied in the chain direction, is a powerful tool of influence on $q$, leading to unusual changes in transport properties of $o$-TaS$_3$ [5-11], such as: different strain-dependences for the Ohmic conductance (with a maximum at a critical strain $\varepsilon_c$) and for the nonlinear one (with a minimum at $\varepsilon_c$); strain-induced decrease of $T_P$ and an increase of $E_\Delta$; disappearing of NBN and an emergence of ultra-coherent CDW near $\varepsilon_c$. The results imply an increase of incommensurability value with a growth of the strain [11], i.e. a growth of solitons concentration. Till now all the strain-induced phenomena in $o$-TaS$_3$ were studied at high temperature range between $T = 66$ K and $T_P$. Here we present the results of the experimental study of the uniaxial strain influence on the low-temperature Ohmic conduction at 10 K $< T <$ 77 K together with a first observation of the strain effect on the photoconduction, which appears at the same temperature region [3].

For the study we have prepared a structure (see insert in Fig. 1) on the base of high-quality $o$-TaS$_3$ crystal ($E_T \sim 0.5$ V/cm, cross section $S \sim 3$ μm$^2$) consisting of three segments: part A – without strain, central buffer part C, part B – with a strain $\varepsilon = \Delta L_B/L_B \approx 1\%$ (where $\Delta L_B$ is a change in a part B length L$_B$), a contact width was $\sim 0.2$ mm. All conductance measurements were done along the chain direction in two-probe configuration in the voltage-controlled regime. IR LED, providing light intensity $W = (10^{-4} - 30)$ mW/cm$^2$ at the sample position, was used; the photon energy $\hbar\omega = 1.3$ eV, optic Peierls gap value $2\Delta_{opt} = 0.25$ eV at $T = 40$ K [12]. The usual AC modulation method (modulation frequency $f = 4.5$ Hz, meander) was used for the photoconduction measurements.

Fig. 1 shows temperature dependences of the Ohmic conductance for the segments A, $G_A(T)$, (upper blue curve) and B, $G_B(T)$, (red curve) together with corresponding sets of temperature dependences of photoconductance, $\delta G_A(T)$ and $\delta G_B(T)$, at different $W$ (all values

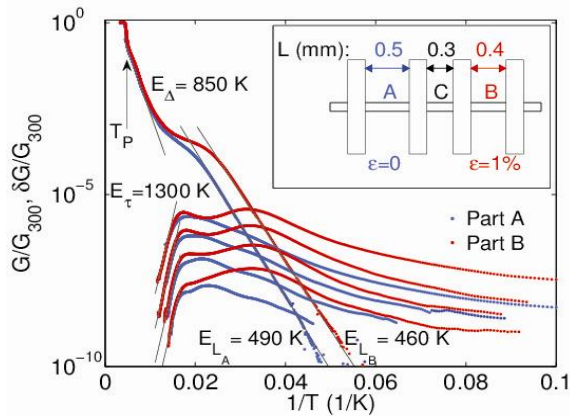 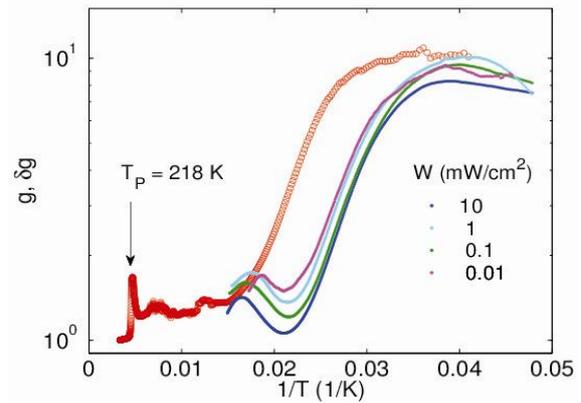

Fig. 1. Temperature dependences of Ohmic conductance $G(T)$ for the segments with and without strain (upper curves) together with corresponding sets of temperature dependences of photoconductance $\delta G(T)$ at following light intensities $W$, top down: 10, 1, 0.1, 0.01 mW/cm$^2$. The insert shows the drawing of the studied structure.

Fig. 2. Temperature dependence of the strain-induced relative change of the Ohmic conductance $g=G_B/G_A(T)$ (red circles) and a set of the similar dependences of the photoconductance change $\delta g=\delta G_B/\delta G_A(T)$ (dots) at different $W$.

$G_A$, $G_B$, $\delta G_A$, $\delta G_B$ are normalized to corresponding room-temperature conductances, $G_{A\_300}$ and $G_{B\_300}$). At high $T$ the stain-induce changes of the dependences are not so dramatic: one can see a smoothing of the Peierls transition, a $T_P$ decrease ~ 6 K and a small (≈ 30 %) $G(T)$ growth, while $E_\Delta$ does not noticeably change for this sample. The activation energy of the photoconductance, $E_\tau$, reflecting temperature dependence of the non-equilibrium current carrier recombination time [3], also does not show a noticeable change under the strain. The low-temperature changes are much more substantial: an additional large contribution to both the conductance and photoconductance (an increase of the main peak and an appearance of a new one) is observed. The value of $E_L$ slightly (≈ 7 %) increases with the strain.

Fig. 2 shows temperature dependences of the strain-induced relative changes of both the conductance $g=G_B/G_A$ and photoconductance $\delta g=\delta G_B/\delta G_A$ (for each $W$). The sharp peak of $g$ at $T_P$ corresponds to suppression of $T_P$ by the strain. Whereas $g$ and $\delta g$ experience a step-like growth at slightly different temperatures, the final low-temperature values of $g$ and $\delta g$ (for all $W$ levels, which differ by 3 orders) being practically the same.

The observed features are consistent with a simple model implying strain-induced increase of concentration of solitons which contribute into both conduction and photoconduction. Further investigations are required to verify this assumption.

The work was supported by RFBR project 14-02-01236.

# XXXVII Совещание по физике низких температур

## Программа и тезисы докладов

## 29 июня – 3 июля 2015 г.